\documentclass[twoside,12pt]{article}
\usepackage{epsfig}

\def\Journal#1#2#3#4{{#1} {#2} (#4) #3 }

\def\NPA{{\em Nucl. Phys.} A}

\def\PLB{{\em Phys. Lett.} B}

\def\PRL{\em Phys. Rev. Lett.}

\def\PRD{{\em Phys. Rev.} D}
\def\PRC{{\em Phys. Rev.} C}

\def\ANNP{\em Ann. Phys.}

\def\JPG{{\em J. Phys.} G}

\newcommand{\be}{\begin{equation}}
\newcommand{\ee}{\end{equation}}
\newcommand{\bea}{\begin{eqnarray}}
\newcommand{\eea}{\end{eqnarray}}

\begin{document}

\title{ \vspace{1cm} Transport coefficients of a unitarized pion gas}
\author{Juan M.\ Torres-Rincon\\
\\
Dept. F\'isica Te\'orica I, Universidad Complutense de Madrid, Spain}
\maketitle
\begin{abstract}
The latest experimental results in relativistic heavy-ion collisions show that the matter there produced
requires transport coefficients because of the important collective properties
found. We review the theoretical calculation of these transport coefficients
in the hadron side at low temperatures by computing them in a gas composed of low energy pions. The interaction
of these pions is taken from an effective chiral theory and further requiring scattering unitarity.
The propagation of $D$ and $D^*$ mesons in the thermalized pion gas is also studied in order to extract the 
heavy quark diffusion coefficients in the system. 
\end{abstract}

\section{Introduction: Transport Coefficients}

Transport coefficients are quantities that describe how the system tends to equilibrium when an external 
perturbation has been applied to it. This perturbation produces a gradient of some hydrodynamical
field (pressure, temperature, velocity...) followed by a response flow in the system (Le Chatelier
principle). This response flow pushes the system towards the equilibrium state. At first order in gradients, the proportionality coefficient
between the hydrodynamical gradient and the response flow is called a transport coefficient (TC):
\be \textrm{TC} \times \nabla \textrm{(Hydrodynamics Field)} = - \textrm{Response Flow} \ . \ee

The response flow corresponds to the motion of a conserved quantity density so there exists one transport coefficient
related to each conserved quantity of the system. Some classical examples of transport coefficients are the shear and bulk viscosities $\eta$, $\zeta$ both
related to the momentum conservation, the heat or thermal conductivity $\kappa$ related to the energy and particle conservation and the electrical conductivity $\sigma$ if
an electric charge is present in the system.

In a relativistic heavy-ion collision, some of them can be extracted from those observables encoding the collective behaviour of the 
plasma. One of the most remarkable transport coefficient extracted in this way is the shear viscosity, usually normalized to the entropy density $\eta/s$.

\section{Flow Harmonics and Viscosities}

 The way to estimate the value of the viscosity over the entropy density is to compare the experimental measurement of a collective observable to
the result given by hydrodynamics simulations (for which $\eta/s$ is an input parameter). The most impressive collective phenomenon is the presence of momentum anisotropies in the 
expanding fireball due to initial spatial anisotropies in noncentral collisions. The quantitative effect of these anisotropies is reflected in the flow harmonics $v_n$, defined
as the averages over the measured particles in some event:
\be v_n = \langle \cos n (\phi_i - \Psi_{RP} ) \rangle \ , \ee	
where $\phi_i$ is the azimuthal angle of the outgoing particles and $\Psi_{RP}$ is the reaction plane, defined by the beam axis and the impact parameter vector between the two incoming nuclei. 
For a recent estimation of the flow harmonics made by the ALICE collaboration see \cite{alice11}.

\begin{figure}[tb]
\begin{center}
\epsfig{file=v2mbGlauber.eps,height=5cm,width=7cm} \ \ \ \ \ \ \ \ \ \
\epsfig{file=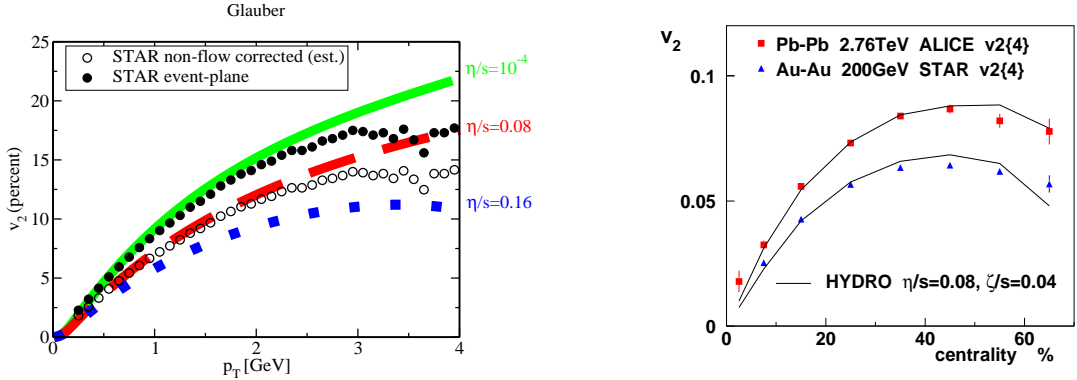,height=5cm,width=6.4cm}
\caption{Left panel: Differential elliptic flow as a function of the transverse momentum as measured at RHIC (dots) and extracted from the hydrodynamic calculation in \cite{luzum08} (lines) for several values of $\eta/s$ and Glauber initial conditions.
Right panel: $p_T$-integrated elliptic flow as a function of the centrality as measured at RHIC and LHC compared to the extracted $v_2$ from the hydrodynamic code of \cite{bozek11} for the optimal values of $\eta/s$ and $\zeta/s$ and Glauber initial conditions. 
Figures courtesy of M. Luzum and P. Romatschke (Copyright 2008 by The American Physical Society) and P. Bozek, respectively. \label{fig:hydro}} 
\end{center}
\end{figure}

For noncentral collisions, the most prominent flow harmonic is the elliptic flow $v_2$ and it is quite dependent on the value of $\eta/s$. By matching the experimental behaviour of $v_2$ and the hydrodynamic calculations
one can estimate the approximate value of $\eta/s$. This value turns out to be very close to $\eta/s \sim 1/(4\pi)$ (see left panel of Fig.\ref{fig:hydro}). The value of the bulk viscosity 
over entropy density ($\zeta/s$) can also be estimated following the same method (see right panel of Fig. \ref{fig:hydro}).

   The newest investigations along these lines are obtaining more precise values for the two viscosities in order to restrict other hydrodynamic parameters and discriminate among different initial condition models.
As an example, the more detailed calculations have began to include the entire temperature dependence of the shear viscosity (see for instance \cite{niemi11}) and of the bulk viscosity (see for example \cite{song08}).
We are going to describe the theoretical calculation of these transport coefficients and its temperature dependence in the low temperature phase, i.e. the hadronic medium.

\section{Hadronized Medium: The Pion Gas}

    We will consider that the matter produced after the hadronization time is mainly composed by pions. Experimentally, it has been found
that the pion multiplicity is of the order of ten times larger than the total multiplicity of the next abundant degree of freedom, the kaon.
At moderate temperatures $T \sim 140$ MeV the next mesonic states (kaons and $\eta$ mesons) should be included in the calculation.

    The properties of a pion gas at low temperatures are not calculable from perturbative QCD. We will
use effective field theory (ChPT) to describe the interaction between pions. The $SU(2)$ chiral perturbation theory (ChPT)
provides a good description of the scattering amplitudes in the energy range we are interested in.

   We will assume that at low temperature the pion gas is dilute enough to only consider binary elastic collisions. 
The suppression of the number changing processes ($2 \to 4$ or $4 \to 2$) comes from two sources. First, the Lagrangian
terms describing these processes are of higher order in momentum and therefore are polynomially suppresed at low temperature (as long
as $p \sim \sqrt{mT}$, where $m$ is the physical pion mass $m=138$ MeV).
Second, in the $2 \to 4$ process the energy needed to create two more pions makes this scattering
suppressed by a Boltzmann factor of $e^{-2m}$. Finally, the process $4 \to 2$ is evidently opposed to the spirit of Boltzmann assumptions
for molecular chaos and does not occur at low density. The encounter of four pions is manifiestly unlike to occurs in a dilute gas. Including only elastic scattering, the pion
number is an effectively conserved quantity and a (pseudo-)chemical potential for the pion must be introduced.


The Lagrangian of ChPT with physical pion masses follows a systematic counting based on terms with 
even powers of pion momentum and pion mass over the scale $\Lambda  \equiv 4\pi F \simeq 1$ GeV:
\be \mathcal{L} = \mathcal{L}_2 + \mathcal{L}_4 + \cdots \ , \ee
where $\mathcal{L}_{2n}$ is $\mathcal{O}[ (p/\Lambda)^{2n}]$ and $\mathcal{O} [ (m/\Lambda)^{2n} ]$. The $G-$parity conservation guarantees that $2n$ is an even number
and it amounts to have interactions between an even number of pions. 

The lowest order contribution to the scattering amplitude is the elastic scattering of two pions. We want to stress again that inelastic channels
are suppresed at Lagrangian level (6-particle interaction is described by $\mathcal{L}_6$) and because of Boltzmann 
suppression in the final phase-space.

From the power counting scheme in the ChPT Lagrangian the partial scattering amplitudes $t_{IJ}$ (labelled by the isospin and angular momentum channels) are
essentially polymonials in momentum:
\be \label{eq:ampli} t_{IJ}^{(2n)} \sim \mathcal{O} (p^{2n}) \ , \ee
where this term comes from the part of the Lagrangian of order $\mathcal{L}_{2n}$. Truncating the expansion at some finite term gives a scattering amplitude that
increases with energy as $t_{IJ}^{(2n)} \sim s^n$. This effect causes an unphysical increase of the pion-pion cross section at moderate energies and leads to a breaking
of the exact unitarity condition for the scattering amplitude. This conditions reads:

\be \label{eq:unit} \textrm{Im } t_{IJ} (s) = \sqrt{1-4m^2/s} \ |t_{IJ} (s)|^2 \ . \ee
The partial amplitudes in (\ref{eq:ampli}) only satisfy this equation perturbatively, e.g. $\textrm{Im } t_{IJ}^{(2)} (s)=0$, $\textrm{Im } t_{IJ}^{(4)} (s) = \sqrt{1-4m^2/s} \  |t_{IJ}^{(2)} (s)|^2$...

We use the Inverse Amplitude Method \cite{dobado89},\cite{dobado96}, a dispersive method that constructs a new partial amplitude $\tilde{t}_{IJ} (s)$ from
the pertubative ones that exactly satisfies the unitarity condition (\ref{eq:unit}). At lowest order, the unitarized amplitude turns out to be:
\be \label{eq:unitar} \tilde{t}_{IJ} (s)= \frac{t_{IJ}^{(2)} (s)}{1- t_{IJ}^{(4)}(s) / t_{IJ}^{(2)} (s)} \ . \ee

This new amplitude, as rational combination of perturbative amplitudes, makes the pion-pion cross section saturate at moderate energies and allows to dynamically generate 
the $\rho$ and $\sigma$ resonances in the $IJ=11$ and $IJ=00$ channels, respectively.


    For calculating the transport coefficients, one needs to solve a transport equation for the one-particle distribution function $f^{\pi}$. This transport equation is of the well-known form:
\be \label{eq:transport} \frac{df^{\pi}}{dt} = C[f^{\pi},f^{\pi}] \ . \ee
For a gas of pions the explicit form of the right-hand side is that of the Boltzmann-Uehling-Uhlenbeck (BUU) equation:
\be \label{eq:BUU} \frac{df^{\pi}_p}{dt} = \frac{g_{\pi}}{2} \int d\Gamma_{12,3p} \left[ f^{\pi}_1 f^{\pi}_2 (1+f^{\pi}_3) (1+f^{\pi}_p) -f^{\pi}_3 f^{\pi}_p (1+f^{\pi}_1) (1+f^{\pi}_2)\right] \ , \ee
where $g_{\pi}=3$ and the measure $d\Gamma_{12,3p}$ constains the details of the pion interaction ($1,2 \rightarrow 3,p$)
\be d\Gamma_{12,3p} \equiv \frac{1}{2E_p} |\overline{T}|^2 \prod_{i=1}^3 \frac{d\mathbf{k}_i}{(2\pi)^3 E_i } (2\pi)^4 \delta^{(4)} (k_1+k_2-k_3-p) \ . \ee
The integral equation (\ref{eq:BUU}) is linearized by using the Chapman-Enskog procedure, consisting in an expansion of the distribution function in powers of the Knudsen number ($\epsilon=$ mean free path / characteristic 
size of the inhomogeneities of the system)
\be f^{\pi}_p = f_{Bose} + \epsilon f^{(1)} + \cdots \ee

In this sense, the perturbative solution is only valid if hydrodynamics can be applied to our system.
Once the transport equation has been linearized an approximate solution for $f^{(1)}$ is obtained expanding this function in powers of the variable $E_p/m$,
\be \label{eq:expansion} f^{(1)} = \sum_{n=0}^N A_n \left( \frac{E_p}{m} \right)^n \ , \ee
up to some finite $N$. Finally, the linearized BUU equation must be numerically inverted to extract the coefficients $A_n$.


Once $f^{(1)}$ is built from the $A_n$ coefficients, the transport coefficient is expressed as a simple integral (or a scalar product):
\be \textrm{TC } = \langle [ \ \nabla(\textrm{Hydrodynamic field} ) \ ]^{-1} | f^{(1)} \rangle \ . \ee

The shape of the hydrodynamic fields and the parametrization of $f^{(1)}$ are dependent on the transport coefficient we are calculating.
The details of the calculation for the shear viscosity and bulk viscosity can be found in \cite{dobado04}-\cite{dobado08} and \cite{dobado11}, respectively.

\begin{figure}[tb]
\begin{center}
\epsfig{file=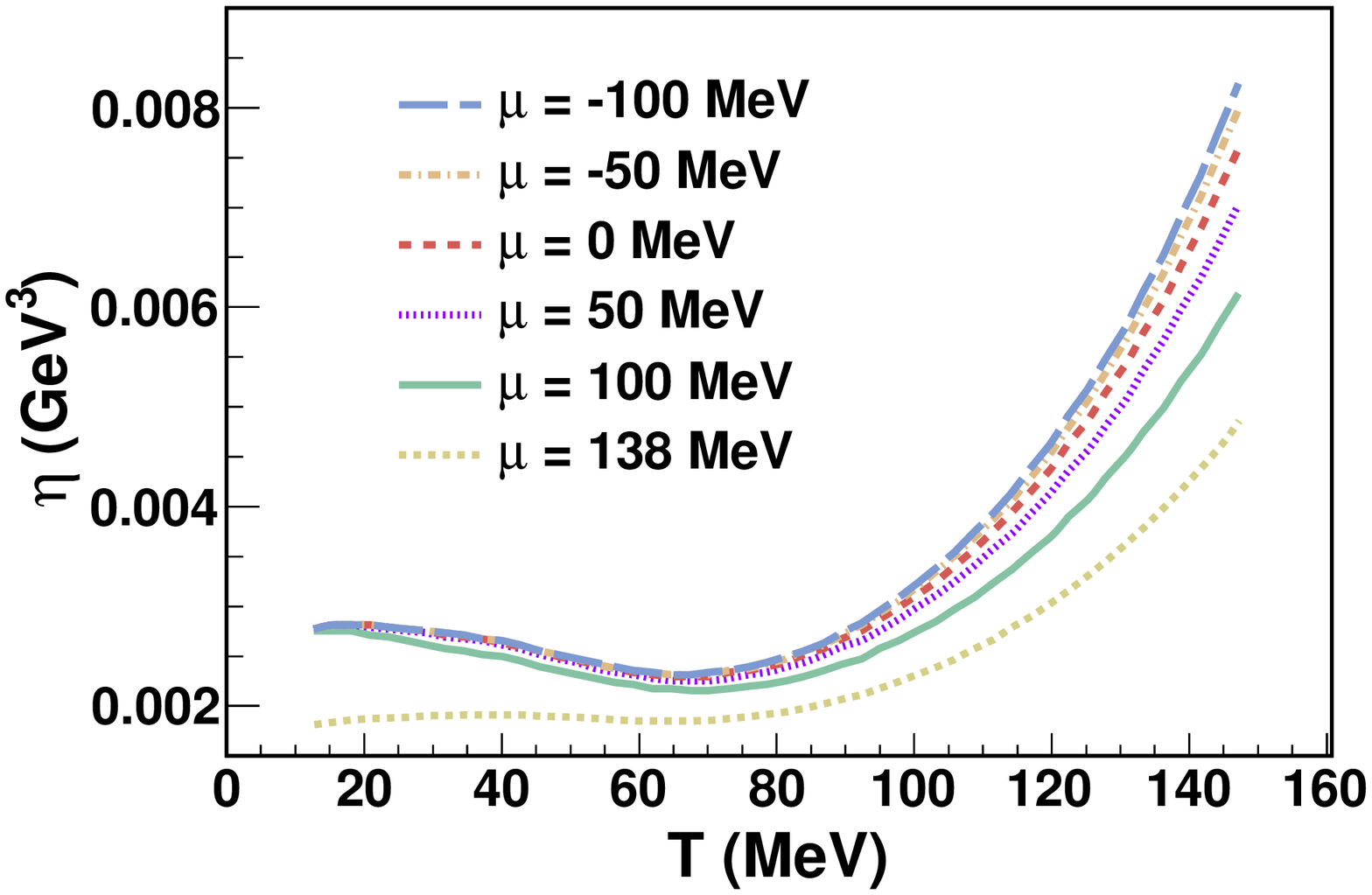,height=4.5cm,width=6cm}
\epsfig{file=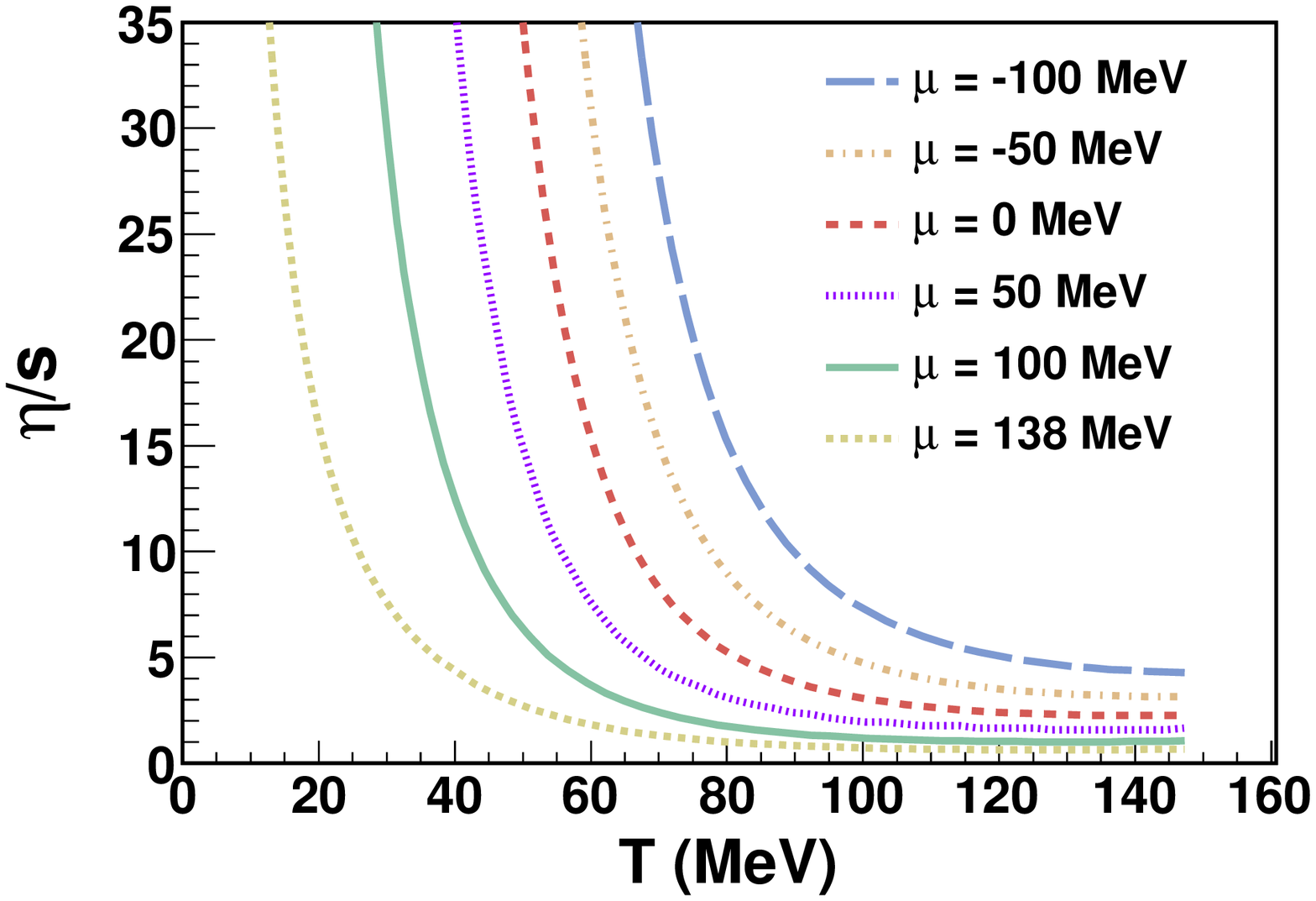,height=4.5cm,width=6cm}
\epsfig{file=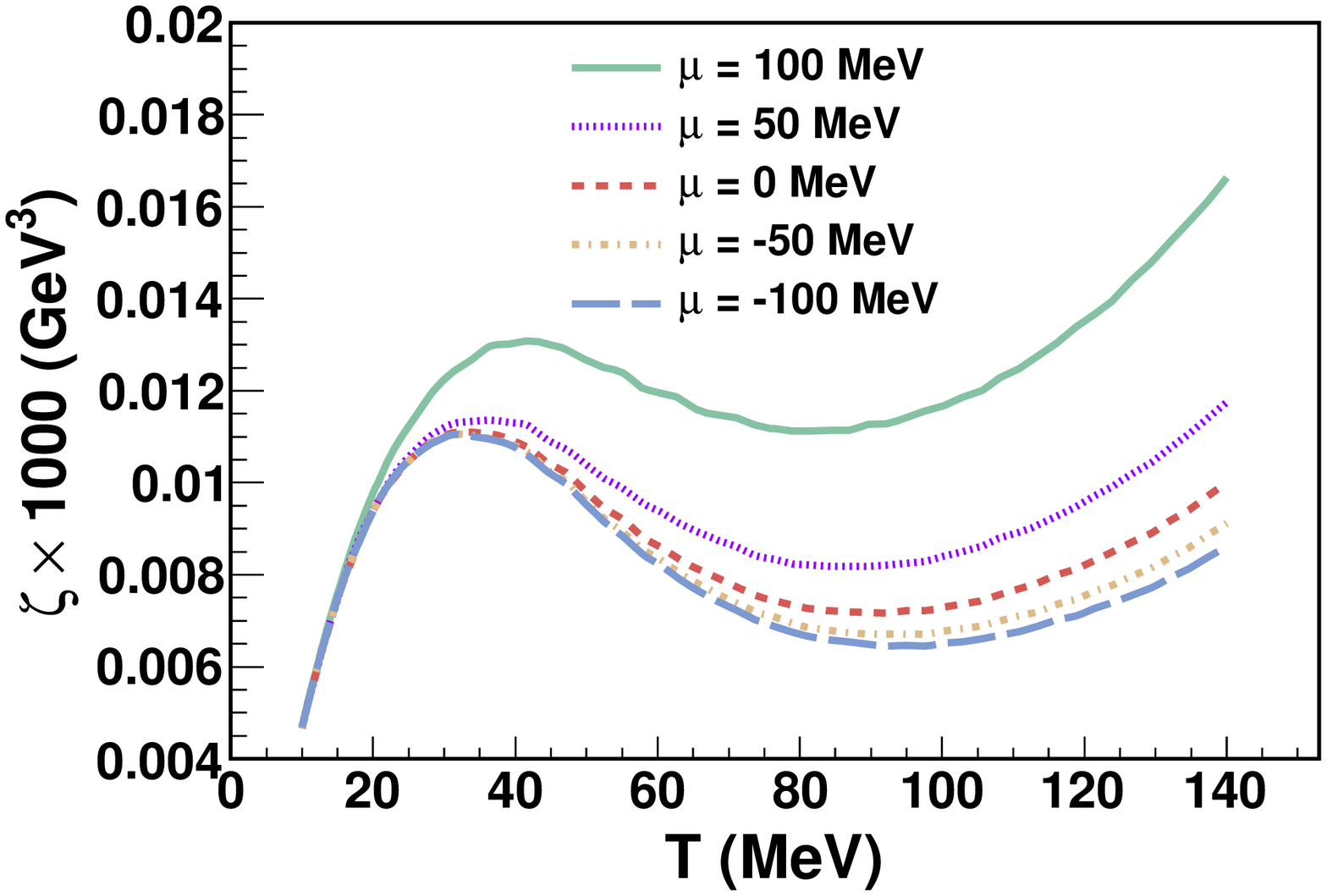,height=4.5cm,width=6cm}
\epsfig{file=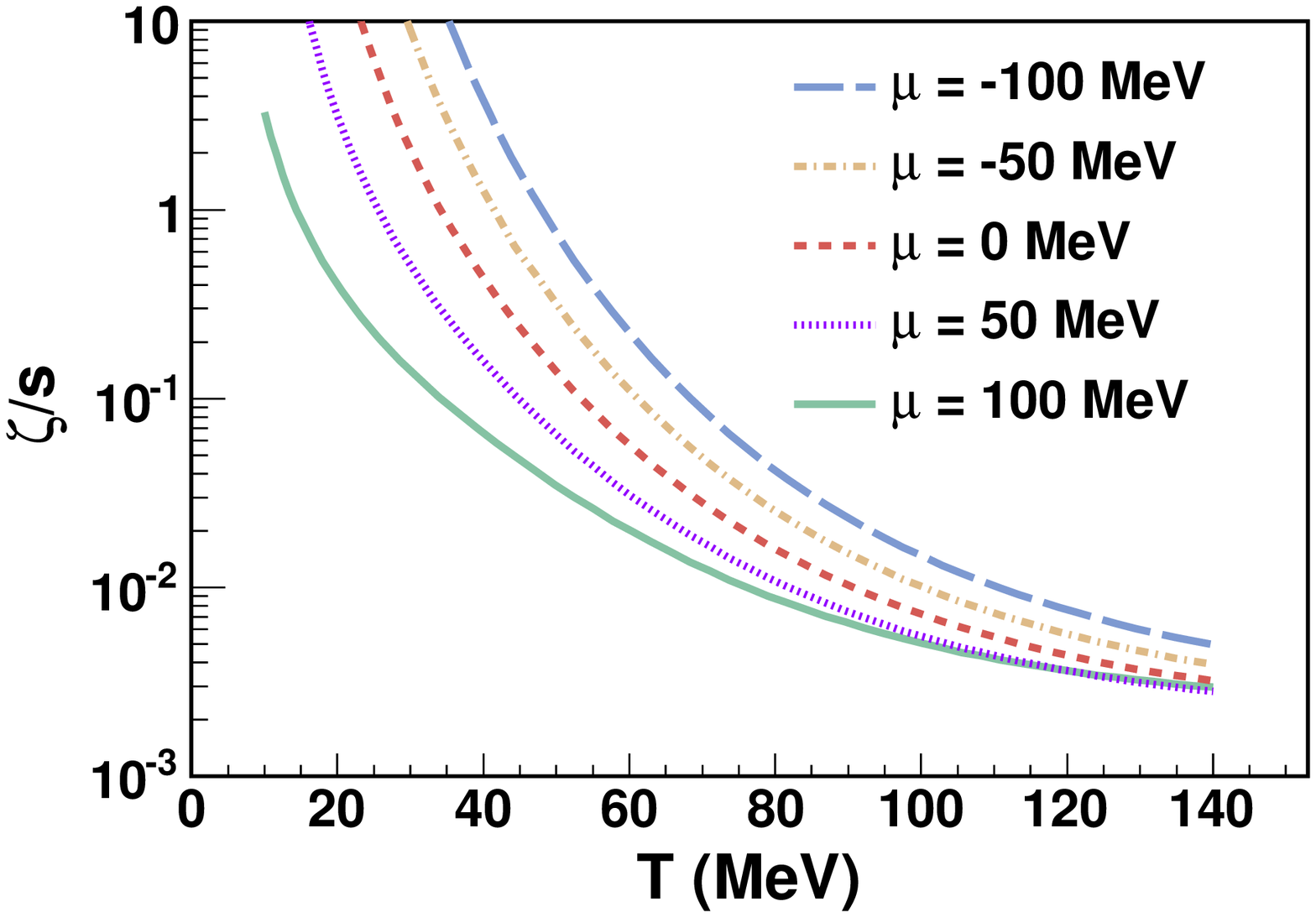,height=4.5cm,width=6cm}
\epsfig{file=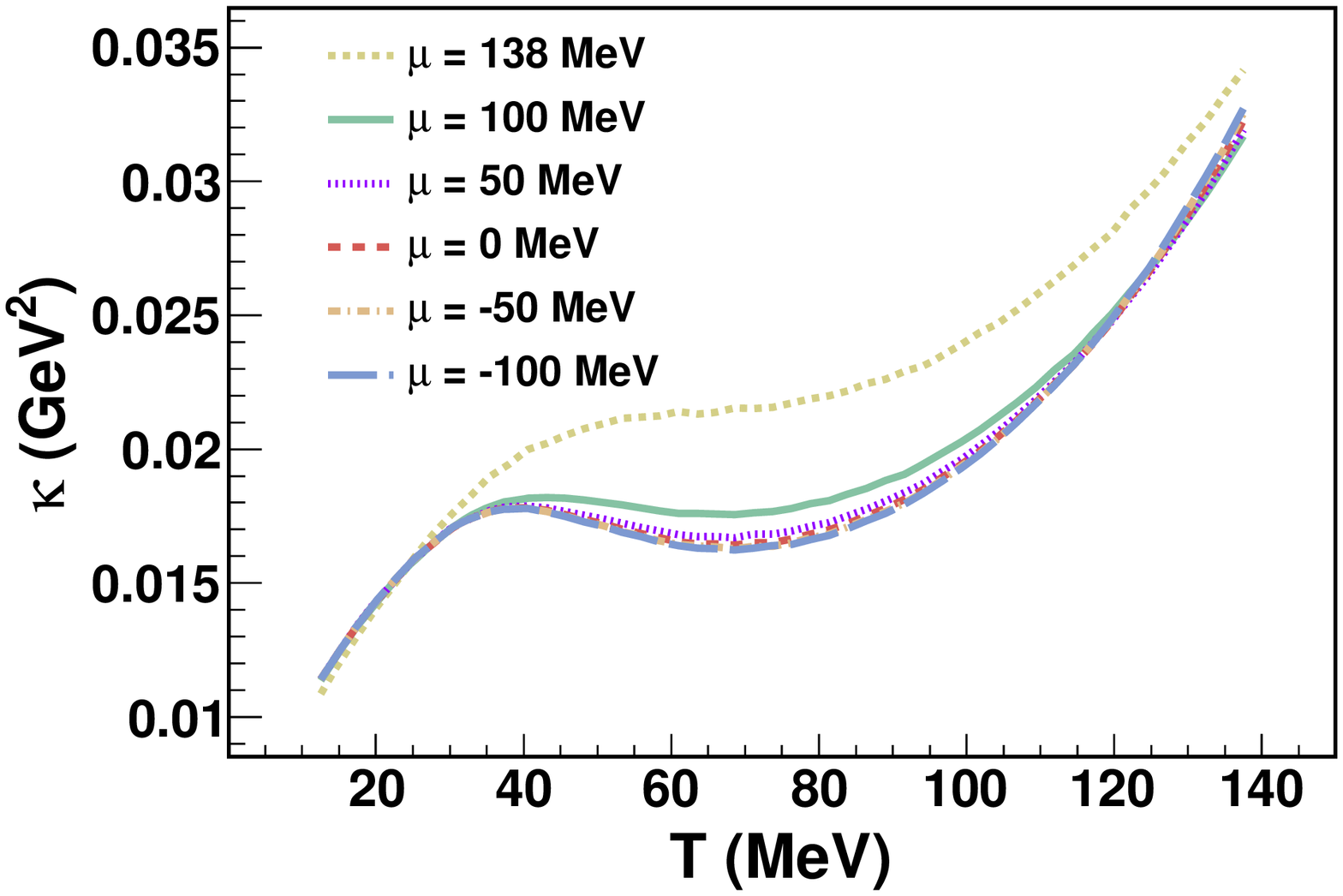,height=4.5cm,width=6cm}
\epsfig{file=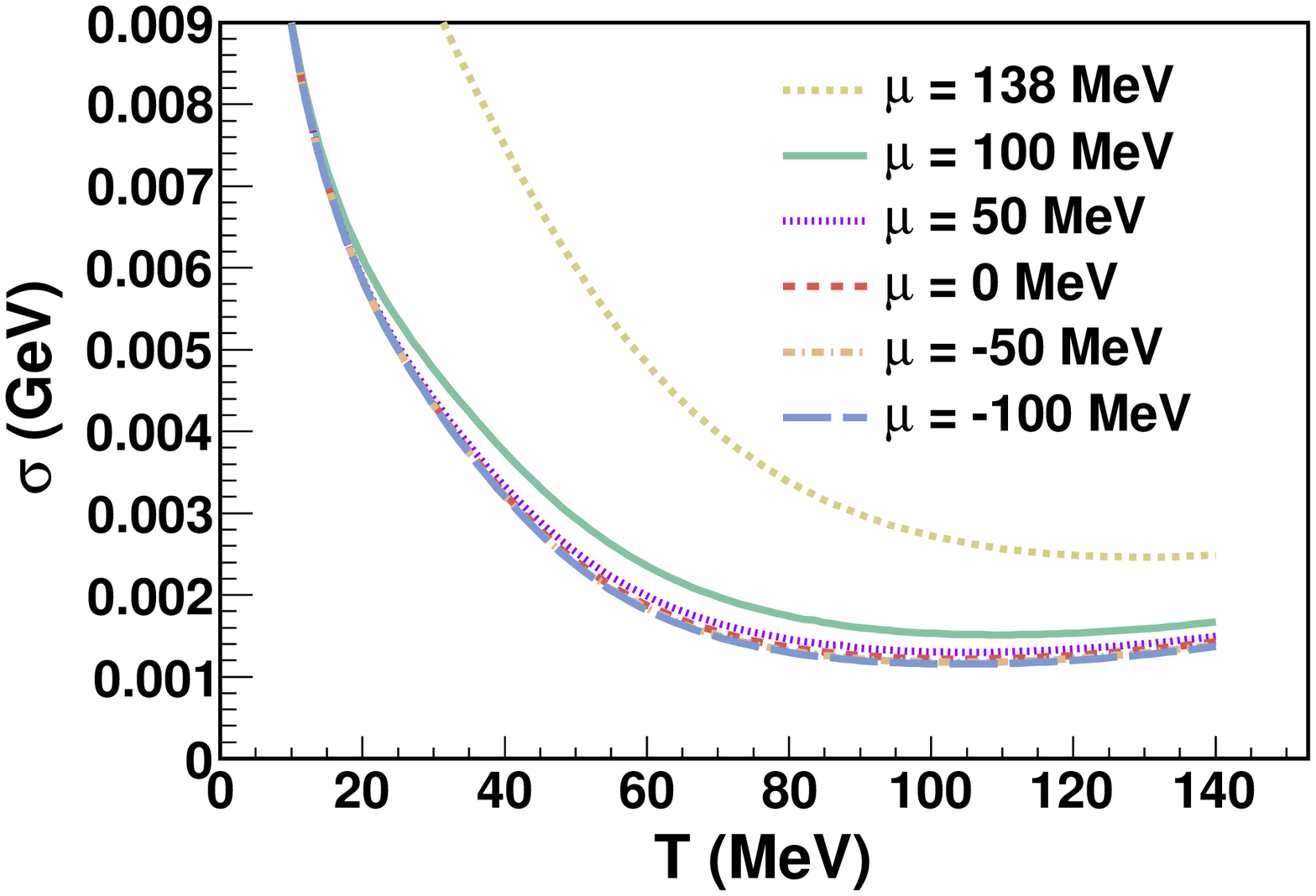,height=4.5cm,width=6cm}
\caption{Transport coefficients of a unitarized pion gas. Top panel: Shear viscosity, shear viscosity over entropy density and scaled bulk viscosity. Bottom panel: Bulk viscosity over entropy density, thermal conductivity and electrical conductivity. \label{fig:coeffs}} 
\end{center}
\end{figure}

 In Fig. \ref{fig:coeffs} we show the results for the transport coefficients of a pion gas as a function of temperature and for several values of the pion chemical potential ($\mu \le m$).
In the top panel we show the shear viscosity, the shear viscosity over entropy density and the bulk viscosity. In the bottom panel we show the bulk viscosity over entropy 
density, the thermal conductivity and the electrical conductivity.

\section{Adding Heavy Quarks to the Medium}

We now include a heavy degree of freedom ($c$ quarks for concreteness) inside the thermalized pion gas. The charm quark 
-hadronized into a $D$ or a $D^*$ meson- propagates in the hot medium and interacts with it transferring energy to the pions due to these
collisions. The charm diffusion coefficients describes the energy loss and momentum spreading of these heavy mesons. 

Experimentally, the diffusion coefficients are important in order to describe the nuclear modification factor, $R_{AA} = \frac{dN_{AA}^e / dp_T}{<N_{coll}> dN_{pp}^e /dp_T}$
and the elliptic flow of the electrons coming from the semileptonic decays of these heavy mesons \cite{phenix08}. For more details we refer the reader to \cite{abreu11}.

 
The interaction of the $D$ and $D^*$ mesons with the pions is described by an effective Lagrangian. This Lagrangian encodes
the chiral and the heavy quark symmetries because of the presence of the pions and the charmed mesons, respectively. These two symmetries
are the guiding principles to construct the sucessive orders of the effective Lagrangian \cite{geng10},\cite{abreu11}.

Most of the terms in the Lagrangian are preceded by low energy constants that are not known. Some of these constants can be fixed by symmetry arguments and others
should be adjusted by asking them to correctly reproduce the experimental decay widths or cross sections of charmed resonances. We work at $NLO$ in the chiral expansion and $LO$ in the heavy quark expansion.
After constructing the perturbative scattering amplitude for the $D-\pi$ and $D^*-\pi$ scatterings, we unitarize the amplitudes in the same spirit as for the pion gas.

The unitarization is done by using the ``on-shell' unitarization technique \cite{oller97},\cite{roca05} that consists in solving the Bethe-Salpeter equation
for the $D-\pi$ rescattering. The final expression of the unitarized amplitude resembles very much the rational structure of (\ref{eq:unitar}). Like the Inverse Amplitude Method, the
``on-shell'' unitarization provides a well-behaved cross section. Moreover, we can dynamically generate the $D_0 (2400)$ and $D_1 (2430)$ resonances, that are automatically incorpored in
the calculation.

 
The transport equation for the distribution function of the $c$ quark reads
\be \frac{df^c}{dt} =C[f^c,f^{\pi}] \ , \ee
in complete analogy with the BUU equation for a gas of pions (\ref{eq:transport}). However the mass of the charmed mesons ($M_D,M_{D^*}$) -much larger
than the pion mass ($m$) or the temperature ($T$)- provides the scale hierarchy
\be \label{eq:masses} M_D, M_{D^*} \ll m \sim T \sim \textrm{transfered momentum} \ , \ee
that simplifies the transport equation for solving the heavy quark distribution function.
The BUU equation tranforms into a Fokker-Planck equation under Eq.~(\ref{eq:masses}):
\be \frac{\partial f^c (t,\mathbf{p})}{\partial t} = \frac{\partial}{\partial p_i} \left[ F_{i} (\mathbf{p}) f^c (t,\mathbf{p})+ \frac{\partial}{\partial p_j} \left( \Gamma_{ij} (\mathbf{p}) f^c (t,\mathbf{p})\right) \right] \ .\ee

Under the assumptions of homogeneity and isotropy one finds three different coefficients that depends on the heavy meson momentum,
one drag force $F(p)$ and two diffusion coefficients $\Gamma_0 (p)$ and $\Gamma_1 (p)$:
\begin{eqnarray}
 F_i (\mathbf{p}) & = & F(p) p_i \ , \\
\Gamma_{ij} (\mathbf{p}) & = &  \Gamma_0(p) \left( \delta_{ij} - \frac{p_i p_j}{p^2} \right) + \Gamma_1 (p) \frac{p_i p_j}{p^2} \ . 
\end{eqnarray}

 Because of the fluctuation-dissipation theorem
there are only two independent coefficients. The explicit expression for these coefficients in terms of the scattering amplitude can be found in \cite{abreu11}.

In the static limit $p\rightarrow 0$ only one of them is independent because the two diffusion coefficients become degenerate and also because the
Einstein formula relates this one to the drag force $\Gamma(p \rightarrow 0)=F(p \rightarrow 0)TM_D$.

There are other related quantities, with direct physical interpretation. On the one hand, the spatial diffusion coefficient $D_x$ -that appears in the description of the Brownian motion- is related
to the static diffusion coefficient. On the other hand, momentum and energy losses are directly related to the drag force. The explicit equations for them are the following:

\be D_x=T^2 /\Gamma; \quad -\frac{dp}{dx}=E \times F(p); \quad - \frac{dE}{dx} = p \times F(p)\ . \ee

\begin{figure}[tb]
\begin{center}
\epsfig{file=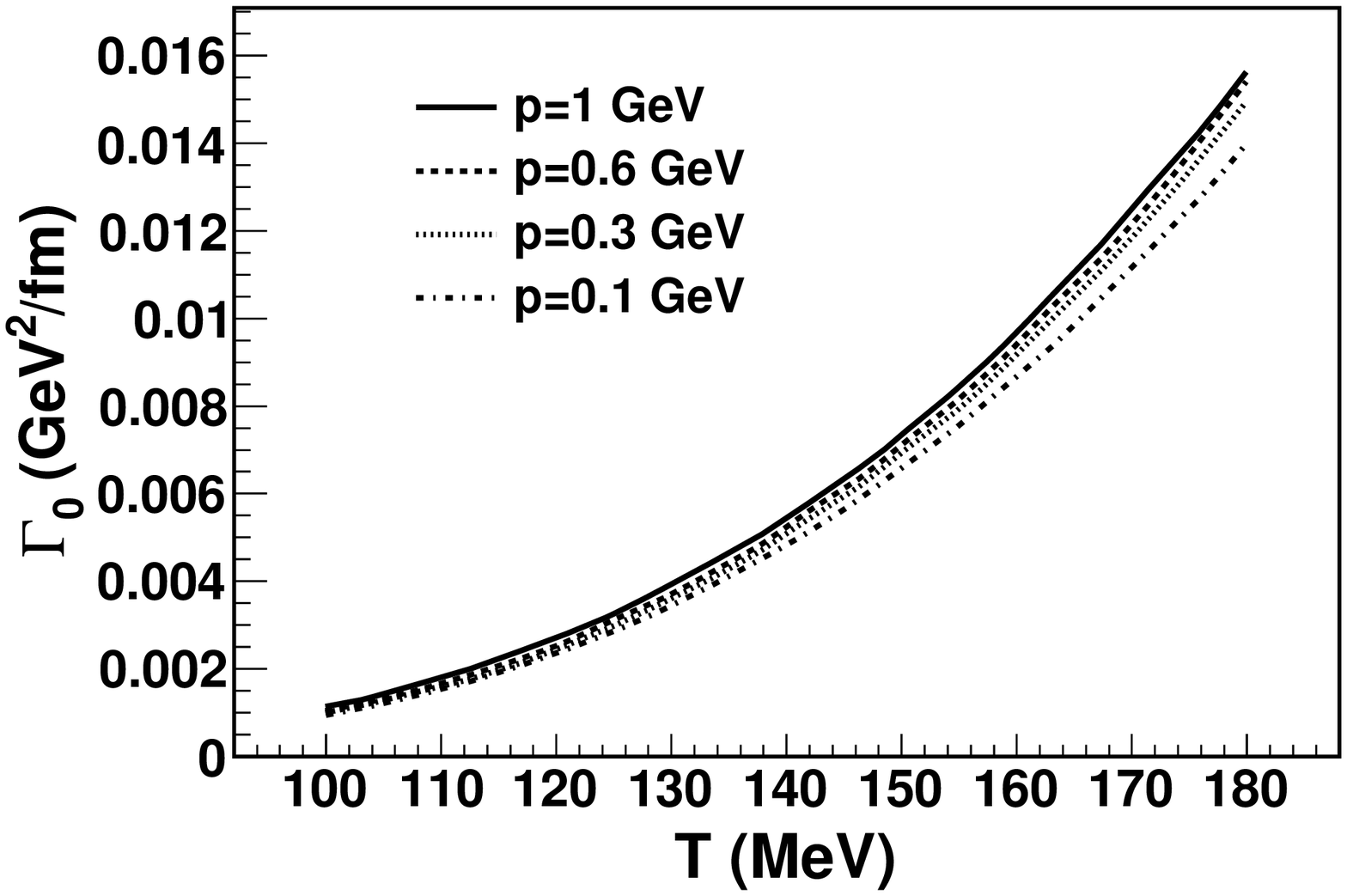,height=4.5cm,width=6cm}
\epsfig{file=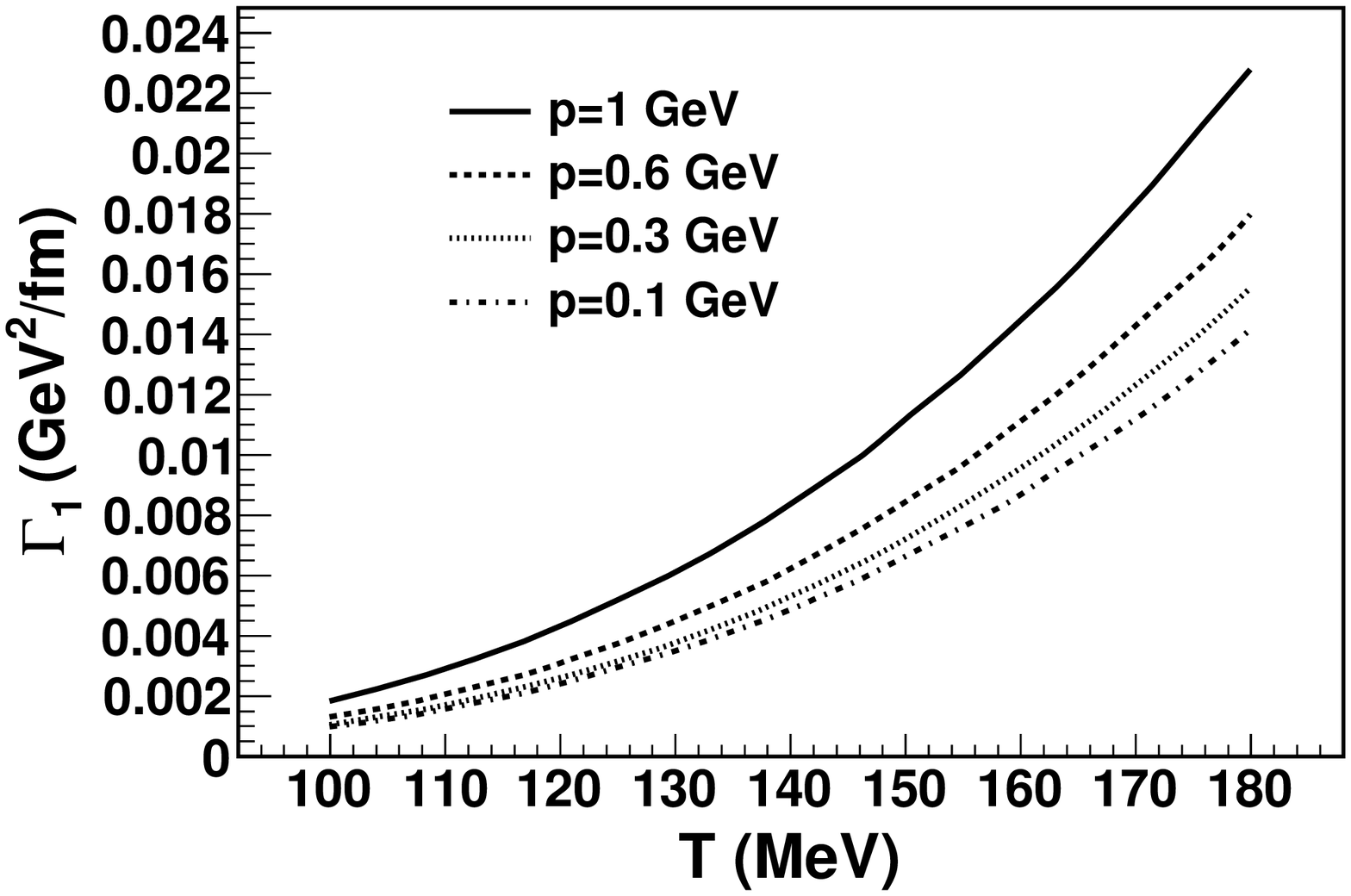,height=4.5cm,width=6cm}
\epsfig{file=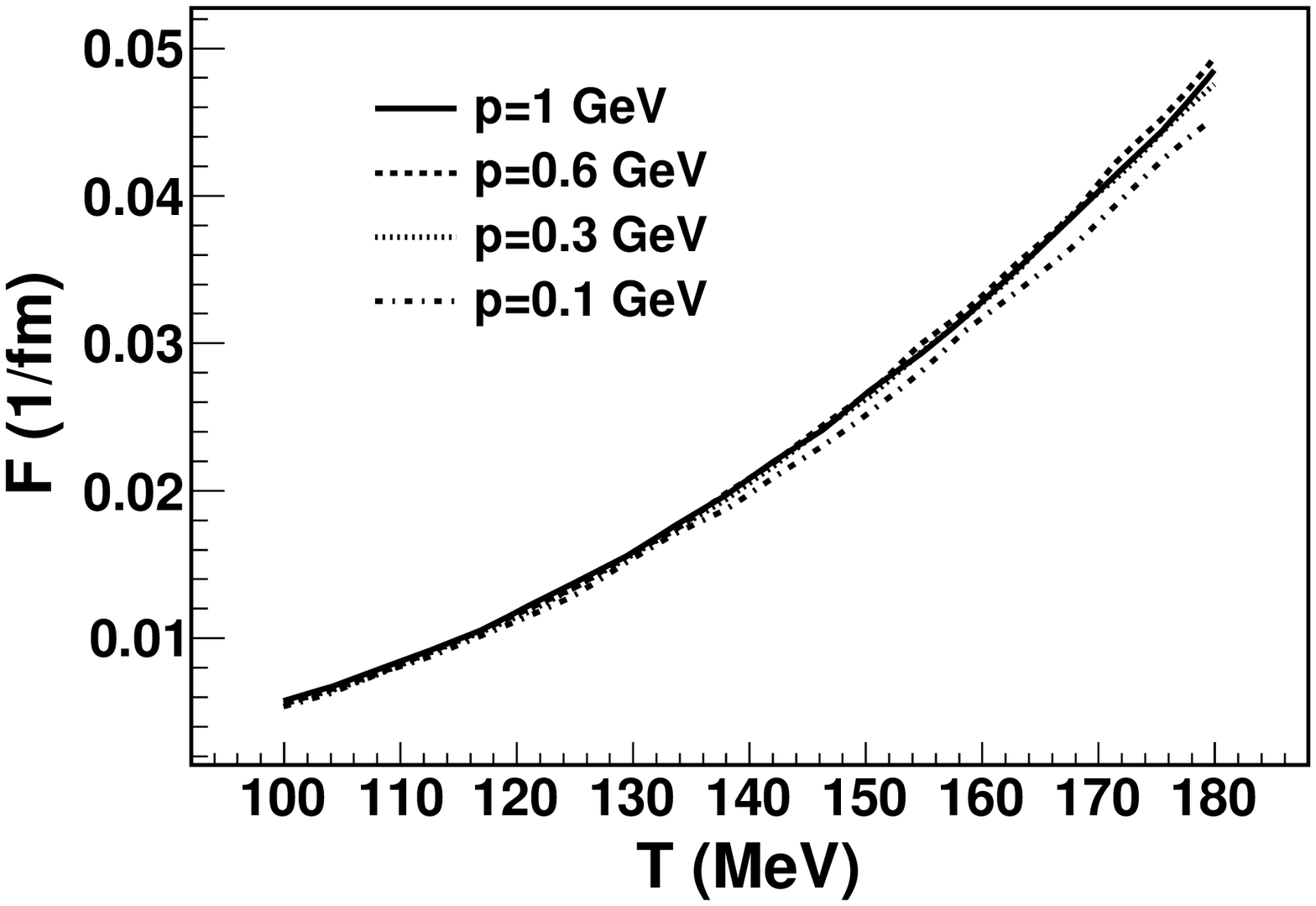,height=4.5cm,width=6cm}
\epsfig{file=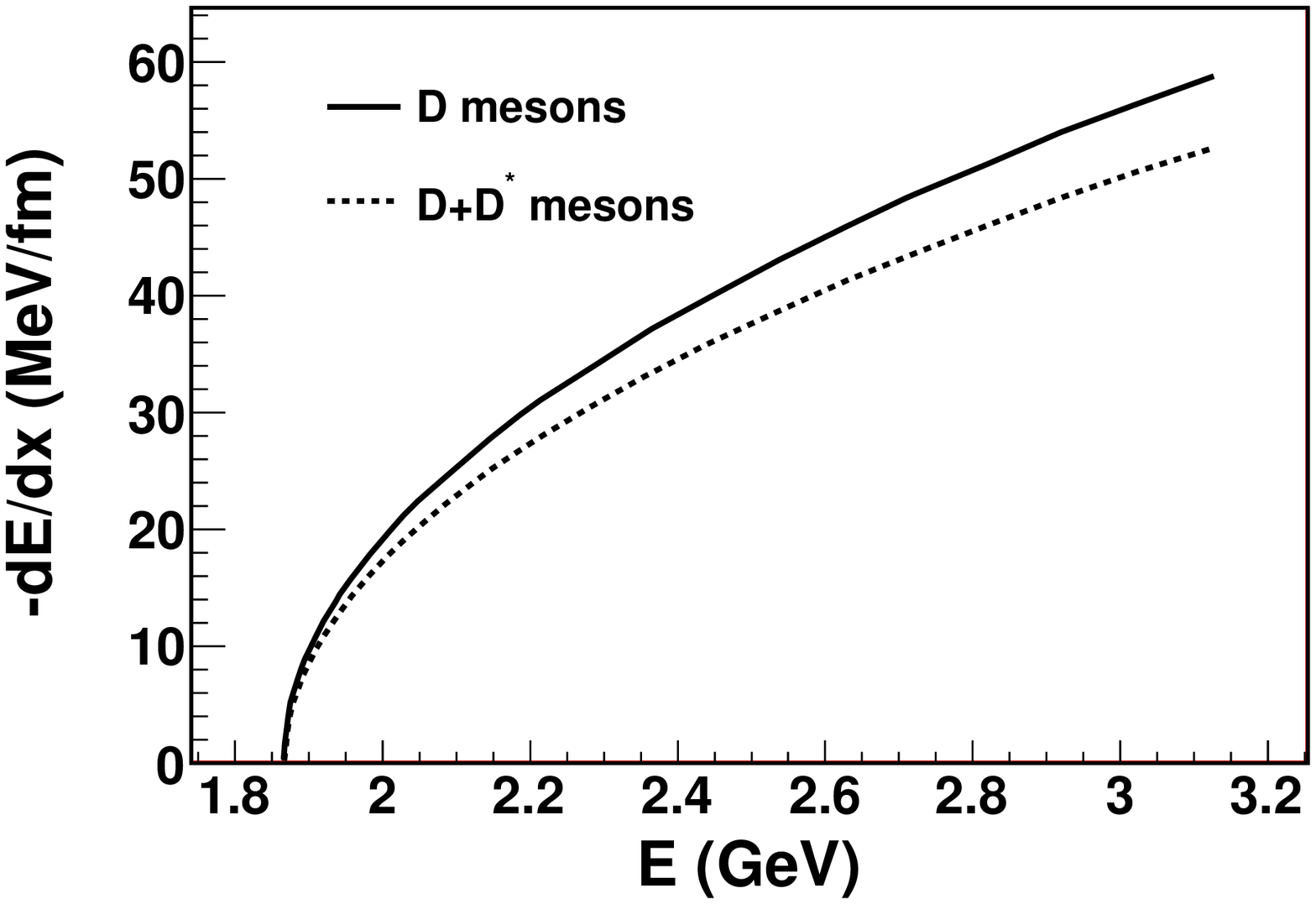,height=4.5cm,width=6cm}
\epsfig{file=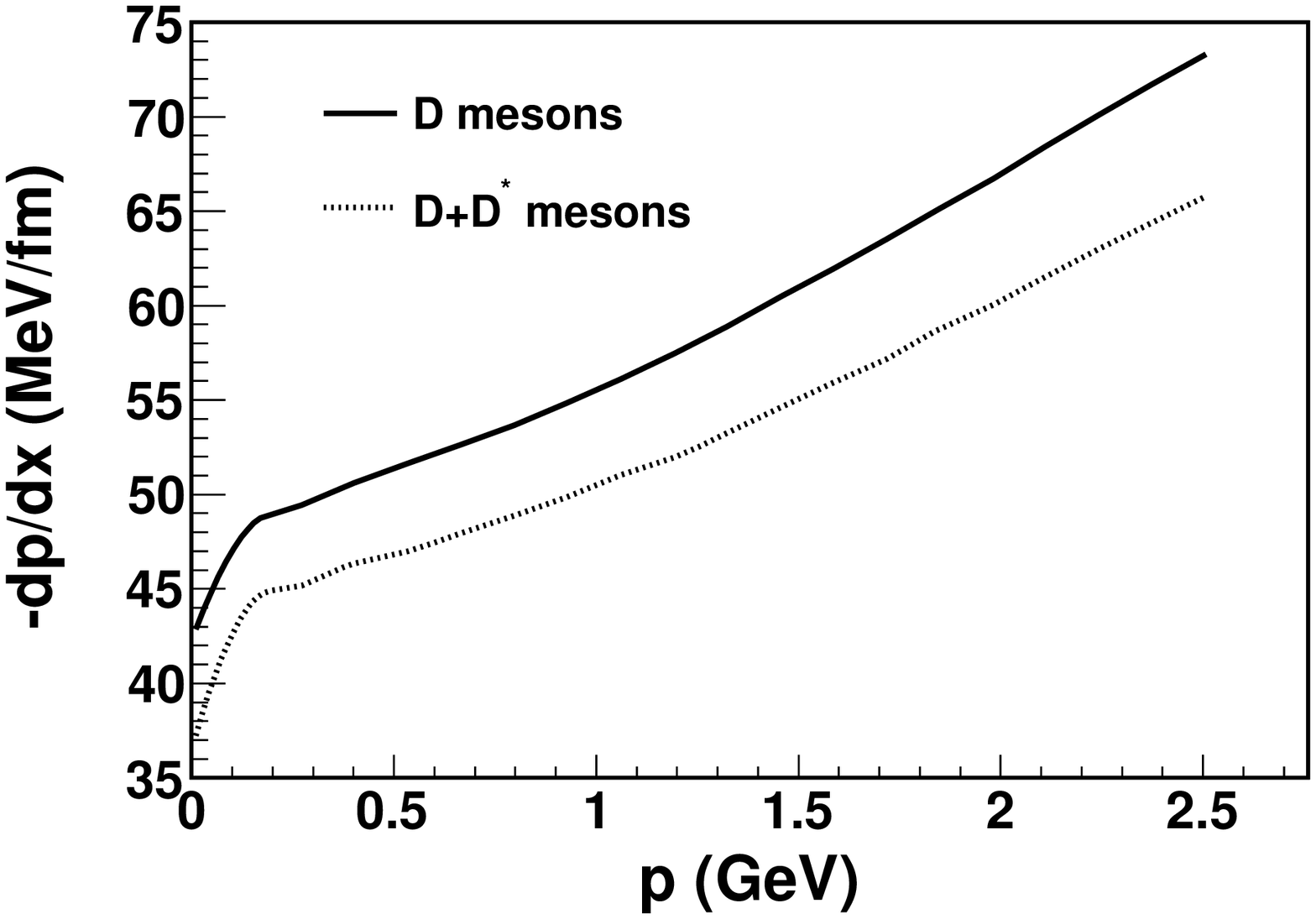,height=4.5cm,width=6cm}
\epsfig{file=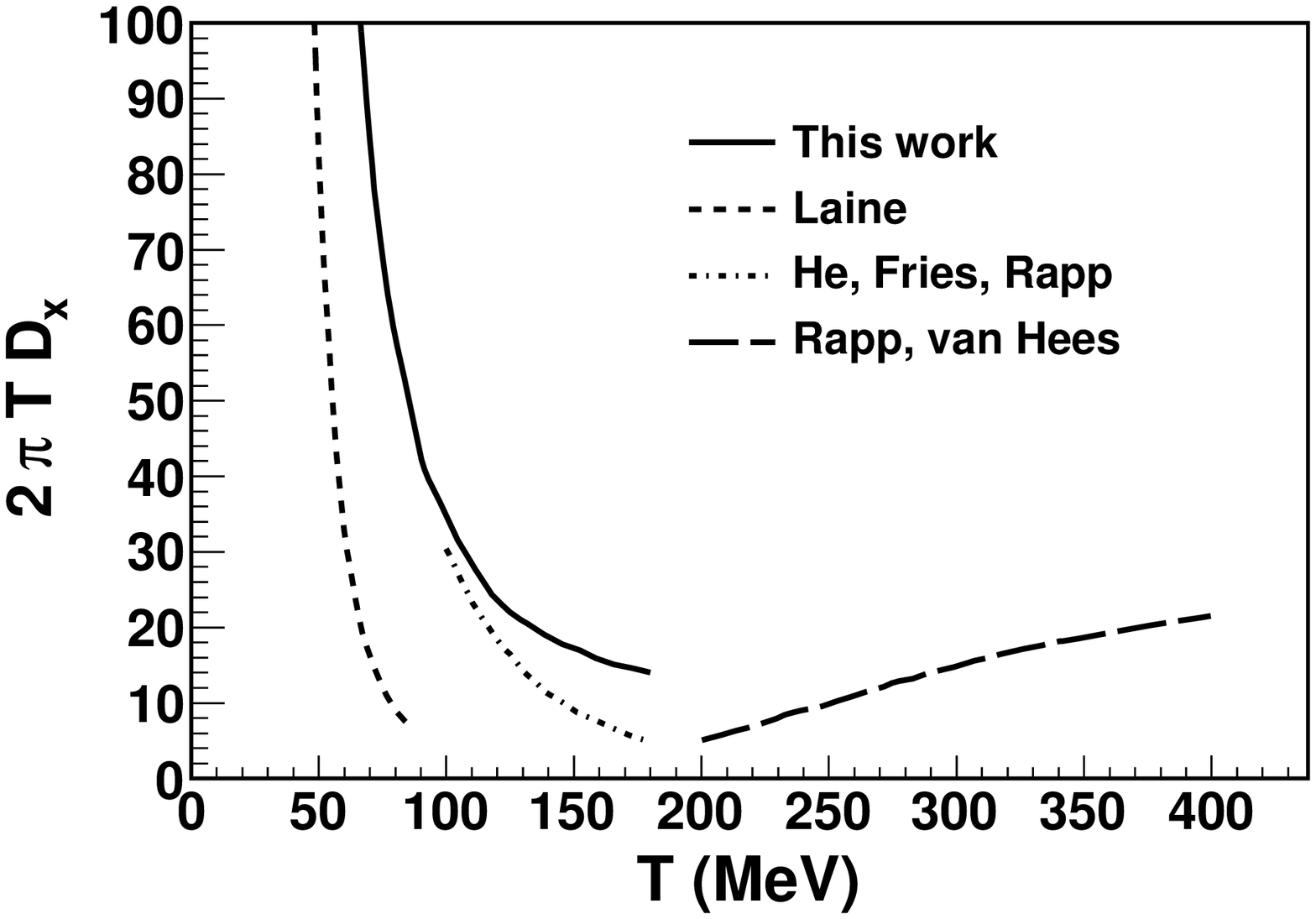,height=4.5cm,width=6cm}
\caption{Top panels: Values of the two diffusion coefficients and the drag force as a function of temperature and momentum. The three coefficients are related through the fluctuation-dissipation theorem.
Bottom panels: Results for the energy loss, momentum loss and the spatial diffusion coefficient in comparison with other approaches. All the plots has been extracted from \cite{abreu11}. \label{fig:diffusion}} 
\end{center}
\end{figure}

We summarize the results in Fig.~\ref{fig:diffusion}. We show the two diffusion coefficients $\Gamma_0 (p)$ and $\Gamma_1 (p)$ as a function of the temperature for selected heavy meson momenta.
These diffusion coefficients enter in the description of the momentum anisotropies, i.e. the elliptic flow of the heavy mesons. We also show the same dependence of the heavy meson drag force, that 
tells us the momentum losses of the heavy meson affecting to the nuclear modification factor of $D$ and $D^*$ mesons. In the same figure, we show the energy and momentum losses of the heavy
 meson that for instance, at $p=1$ GeV leaves about $50$ MeV of momentum per Fermi travelled. Finally, we show the spatial diffusion coefficient
$D_x$ together with some other results along the same lines and the calculation for a perturbative heavy quark inside the quark-gluon plasma (see \cite{abreu11} and references therein).

\section{Conclusions}
 
We have described the theoretical calculation of the transport coefficient of a meson gas at low temperatures. We have studied the pure pion gas and the case where charm degrees of freedom are added to this gas.
For both cases, we have used an effective Lagrangian to describe the interactions between the mesons and we have provided unitarization schemes to ensure that the scattering amplitudes are properly unitarized. 
All these coefficients provide valuable information about the experimental observables that reflects collective phenomena of the medium, such as flow harmonics and nuclear modification factors.
We believe that a better understanding of these non-equilibrium effects will provide a better estimation of these observables and an accurate description of the properties of the matter produced at relativistic heavy ion collisions. \\
{ \it Work supported by grants FPA 2008-00592 and AIC10-D-000582 (Spain). The author is recipient of an FPU scholarship from the Spanish Ministry of Education (AP2008-0083).
Work done in collaboration with A. Dobado, F.J. Llanes-Estrada, L.M. Abreu and D. Cabrera.}

\end{document}